\documentclass[10pt,a4paper]{article}
\usepackage[utf8]{inputenc}
\usepackage{amsmath}
\usepackage{amsfonts}
\usepackage{amssymb}
\newcommand{\nuktilde}{\tilde{\nu}_{K}}
\newcommand{\nuk}{{\nu}_{K}}
\usepackage{graphicx}
\usepackage{caption}
\newcommand\blfootnote[1]{%
  \begingroup
  \renewcommand\thefootnote{}\footnote{#1}%
  \addtocounter{footnote}{-1}%
  \endgroup
}
\numberwithin{equation}{section}
\begin{document}

\begin{center}
{
\bf \large GRAPHENE ELECTRONIC STRUCTURE IN \\
CHARGE DENSITY WAVES
}
\end{center}

\begin{center}
{\bf John M. Vail$^{1*}$}\blfootnote{Corresponding author: vail@umanitoba.ca}, O. J. Hernandez$^2$, M. S. Si$^3$ and Z. Wang$^4$
\end{center}

\begin{center}
{ \it
$^1$Department of Physics and Astronomy and Winnipeg Institute for Theoretical Physics,
University of Manitoba,Winnipeg, MB R3T 2N2, Canada
}
\end{center}

\begin{center}
{ \it
$^2$Department of Physics and Astronomy, University of British Columbia,
 Vancouver, BC V6T 1Z4, Canada,and
TRIUMF, 4004 Wesbrook Mall, Vancouver, BC V6T 2A3, Canada
}
\end{center}

\begin{center}
{ \it
$^3$Key Laboratory for Magnetism and Magnetic Materials, Lanzhou University,
Lanzhou, 730000, People’s Republic of China
}
\end{center}

\begin{center}
{ \it
$^4$Department of Physics, College of Science, South China Agricultural University, 
Guangzhou, 510642, People’s Republic of China
}
\end{center}

\noindent
{\bf Abstract.}  We introduce the idea that the electronic band structure of a charge density wave system may mimic the electronic structure of graphene. In that case a class of materials quite different from graphene might be opened up to exploit graphene's remarkable electronic properties. The theory of such materials, along with superconductivity, is based on the material's dynamical, rather than its static, properties. The charge density wave system turns out to have a number of requirements: (1) a specific wave geometry simply related to graphene; (2) a self-consistency among the electrons that requires the net Coulomb and phonon-mediated parts of the electron-electron interactions to be attractive. We develop a model that leads to an analytical expression for the total energy in terms of the effective electron mass $\mu$, the electron density $\rho_0$, and the strength $\nuktilde$ of the net electron-electron interaction. For constant $\mu>0$ and $\rho_0$, we examine the limitations set upon $\nuktilde$ by self-consistency, stability, and the approximation in the electronic state calculation, and find them to be mutually compatible. This demonstrates the viability of our model.

\section{Introduction}
\indent
\indent A simplified model of graphene consists of a planar hexagonal array of carbon atoms (C). Its electronic band structure was first determined by Wallace in a 1947 study of graphite.~[1] It was first isolated and studied in detail by Novoselov et al. in 2004.~[2] Since then, it has been intensively studied, particularly for its potential value as a technological material. It has been found to have a number of extraordinary properties, including high electronic conductivity and enormous strength.\\
\indent The ground state electronic structure of graphene consists of a filled hexagonal first Brillouin zone (BZI), with an energy gap that goes monotonically to zero at the zone corners, from a maximum at the centers of the zone edges. This electronic configuration arises from the particular relationship between the number of conduction (non-core) electrons per atom and the hexagonal atomic ordering.\\
\indent A simple model of the non-core electrons in graphene has a two-dimensional hamiltonian consisting of electron kinetic energy, electron-core (EC) interactions and electron-electron (EE) interactions. Two extreme approximations from this formulation consist of the limiting cases: (1) where (EE) interactions are negligible compared to (EC) interactions, and (2) where (EC) are negligible compared to (EE). Wallace’s treatment for the band structure falls within case (1). Case (2) is representative of models for superconductivity (SC) and for charge density waves (CDW). The theoretical basis for the latter was established by Fröhlich in 1954.~[3]\\
\indent For the cases (SC) and (CDW), a number of characteristics is required of the (EE) interactions. First, they involve pairwise interactions (at least). The problem of solving the Schrödinger equation then includes a self-consistency requirement among all the non-core electrons. Second, the (EE) is not representable in terms solely of pairwise Coulomb repulsions. One needs to include the effect of the phonon field of the crystal upon the non-core electrons, in simplest terms as an attractive pairwise contribution. Not only that, but the combination of the Coulomb repulsion and phonon-mediated attraction needs to constitute a net attraction, at least for some portion of the phonon spectrum.\\
\indent Superconductivity and charge density waves have not been observed in pure, perfect single-sheet graphene, as far as we know. We feel that the CDW being introduced in this work will not be found there, given that the characteristic graphene band structure is correctly modeled by Wallace while neglecting electron-electron interactions. They have been induced, however, in impurity-doped or otherwise modified graphene systems.~[4],~[5]\\
\indent In the present work, we are not studying graphene. Rather, we are considering the possibility that other planar atom-thick systems may exist or be fabricable, such that they sustain CDW, and that the electrons in those systems see a potential due to the periodic CDW which mimics the potential seen in graphene due to the period array of the atomic cores of the crystal.\\	
\indent Our purpose is to determine the material requirements for such a system. In terms of our simple model, there are three physical parameters that may be manipulated experimentally: the effective electron mass $\mu$ in a two-dimensional system, and the dominant Fourier coefficient $\nuk$ in the periodic CDW, and its wave number $K$. We also address the question whether the CDW state is a stable ground state, relative to an approximately uniform-density state. The work elucidates the conceptual and material considerations that arise in creating specific band structures by engineering charge density waves. The outline of the paper is as follows: In Sec.~2, we specify the model that will be analyzed. In Sec.~3, we give the method whereby specific results will be achieved through relatively simple calculation. In Sec.~4, the self-consistency condition (which arises whenever many-body interactions are included) is determined explicitly for our model and method, and its physical meaning is explored. In Sec.~5, the total energy of the CDW system is evaluated in terms of the physical parameters $\mu$, $\rho_0$ and $\nuk$. In Sec.~6, the total energy is evaluated for the uniform-density state of the system, and in Sec.~7, the stability condition for the CDW state is displayed. In Sec.~8, we summarize our results and present our conclusions.

\section{Model: A CDW Having the Graphene Electronic Structure}
\indent
\indent We shall model a planar crystalline system with a number $n$ of non-core conduction electrons in an area $\Omega$. We shall assume that these electrons will form a CDW ground state that mimics the electronic band structure of graphene. In such a CDW state, the electronic density will have an array of minima and maxima whose positions mimic the carbon atomic core sites of graphene at the corner of hexagons, and the interstitial sites at the hexagon centers.

\subsection{\it formulation}
\indent 
\indent We begin with a two-dimensional crystal consisting of atomic cores (not necessarily carbon) and conduction electrons. This contains the implicit assumption that the problem is separable into planar $(x, y)$ variables and a transverse variable $z$ that is omitted. Consider the model hamiltonian:

\begin{equation}
H_n = \sum_{j=1}^{n}\left[ -\frac{\hbar^{2}}{2m}\nabla^{2}_{j}+\sum_{J}\nu_{\text{core}}(\vec{R}_{J}-\vec{r}_{j}) \right] +\frac{1}{2}\sum_{j,j'}^{n}\nu\left(\vec{r}_{j}-\vec{r}_{j'} \right),
\end{equation}
\noindent
where $j$ = $1,2,…n$ are electrons whose two-dimensional position and spin variables are combined in $\vec{r}_j$, $J$ labels atomic cores, $\nu_{\text{core}}$ is electron-core interaction, and $\nu$, the effective electron-electron interaction, is taken to be pairwise and cylindrically symmetrical.\\
\indent Now, the core terms $\left( \sum_{J} \right)$ constitute the periodic potential of the crystal. If the conduction electrons are approximately like non-interacting electrons at the bottom of the first conduction band of the crystal, then the periodic potential may be replaced by an effective band mass $m_b$ in place of the free-electron mass $m$. We express this in terms of a dimensionless parameter $\mu$:
\begin{equation}
m_b = \mu . m.
\end{equation}
We can now rewrite the Hamiltonian $H_n$ in bohr-hartree atomic units:
\begin{equation}
H_{n} = \sum_{j=1}^{n}\left[ -\frac{1}{2\mu}\nabla^{2}_{j}+\frac{1}{2}\sum_{j'=1}^{n}\nu(\vec{r}_j-\vec{r}_{j'})\right] 
\end{equation}
Units of length are now bohr ($a_0$), and units of energy are hartree (Hy), where:
\begin{align}
a_0 = \frac{4\pi \varepsilon_{0}\hbar^2}{m e^2} = 0.529 \text{ A}, & & 1 \text{ Hy} = \frac{\hbar^2}{m a^2_{0}} = 27.2 \text{ eV}.
\end{align}
The pairwise interaction is now expressed in Hy units. Note especially that in eq.~(2.3), the periodic potential, if any, will arise from $\nu(\vec{r})$, expressing the charge density wave, \underline{not} from the periodicity of the atomic ordering in the crystal, which is expressed by the band mass parameter $\mu$.\\
\indent The physics of our model system should include the phonons of the crystal and their energy, and the electron-phonon interaction and its energy. We assume that the electron-phonon interaction produces a modification of the electron-electron interaction and of the phonon-phonon interaction with the resultant effective electron and phonon dynamical variables becoming separable, the electron-phonon interaction having been subsumed into the new electronic and phononic parts of the system Hamiltonian. We can then analyze the electronic part, as in eq.~(2.1), and treat the phonon part as a fixed, constant part of the total energy. In this case, the effective pairwise electron-electron interaction, $\nu(\vec{r})$ in eq.~(2.1), contains a phonon-mediated contribution, in addition to the Coulomb repulsion. While these postulates are not likely to be justifiable in general, the resultant model may mimic specific properties of some systems. 

\subsection{\it the total energy}
\indent
\indent It is our purpose, both in the system’s model and in the computation of its properties, especially its total energy, to use as simplified an approach as possible. The total energy $E_n$ from eq.~(2.3) is:
\begin{equation}
E_n = \langle \Psi_n | H_n | \Psi _n \rangle,
\end{equation}
where $\Psi_n$ is the ground state many-electron wave function and $E_n$ is its eigenvalue. For simplicity, consider the Hartree approximation, where 
\begin{equation}
\Psi_n(\vec{r}_1, \vec{r}_2, ..., \vec{r}_n ) = \prod_{j=1}^{n} \psi_{j}(\vec{r}_j)
\end{equation}
where $\psi_{j}(\vec{r}_j)$ is a set of $n$ orthonormal single particle functions. If such a set is to minimize $E_n$, it must satisfy the Hartree equation:
\begin{equation}
h(\vec{r}).\psi_{j}(\vec{r}) = \varepsilon_{j}.\psi_{j}(\vec{r}),
\end{equation}
where the Hartree operator $h$ is:
\begin{align}
h(\vec{r}) &=  -\frac{1}{2\mu}\nabla^2+\sum_{j'=1}^{n} \langle j' |\nu | j' \rangle ;\\
\langle j' |\nu | j' \rangle &= \int d^2 r'. \psi^{*}_{j'}(\vec{r}').\nu(\vec{r}-\vec{r}').\psi_{j'}(\vec{r}').
\end{align}
From eqs.~(2.7) and (2.8), if we sum over $j$ from $j=1$ to $j=n$, and compare the result that comes from eqs.~(2.3) and (2.6), we arrive at the well-known conclusion that the two-particle contribution to the energy is double counted in the Hartree equation. We shall therefore evaluate the total energy $E_n$ by summing over the $n$ single-particle Hartree energies $\varepsilon_j$, and then subtracting half of the potential energies that are included in that sum:
\begin{equation}
E_n =  \sum_{j=1}^{n}\varepsilon_{j} - \frac{1}{2}\sum_{j,j'=1}^{n}\langle j,j'|\nu|j,j' \rangle ;
\end{equation}
see eqs.~(2.5)-(2.9).

\subsection{\it electron density}
\indent
\indent For our study of CDW, it is useful to include the electron density operator $\rho_{op}(\vec{r})$  in the formulation. Consider:

\begin{equation}
\rho_{op}(\vec{r}) = \sum_{j=1}^{n}\delta(\vec{r}-\vec{r}_j)
\end{equation}
\noindent
We now introduce $\Omega$ as the Born-von Karmann area of a system consisting of $n$ electrons in $N$ primitive unit cells of the Bravais lattice, subject to periodic boundary conditions. In terms of its Fourier analysis, $\rho_{op}(\vec{r})$ is:

\begin{equation}
\rho_{op}(\vec{r}) = \sum_{\vec{k}} \rho_{\vec{k}}.\text{exp}(i\vec{k}.\vec{r})
\end{equation}

\begin{equation}
 \rho_{\vec{k}} = \frac{1}{\Omega} \int d^{2}r' . \rho_{op}(\vec{r}').\text{exp}(-i\vec{k}.\vec{r}') = \frac{1}{\Omega}\sum_{j=1}^{n}\text{exp}(-i\vec{k}.\vec{r}_j).
\end{equation}

\subsection{\it the self-consistent field}
\indent
\indent The two-particle term in the Hartree equation~(2.9) is called the self-consistent field, denoted scf. It is related to the electron density as follows. Consider the Fourier transform relation:
\begin{align}
\nu(\vec{r}) &= \sum_{\vec{k}}\nu_{\vec{k}}.\text{exp}(i\vec{k}.\vec{r}) ;\\
\nu_{\vec{k}} &= \frac{1}{\Omega} \int d^{2}r'.\nu(\vec{r}').\text{exp}(-i\vec{k}.\vec{r}).
\end{align}
Then in the scf we have:
\begin{align}
\sum_{j'=1}^{n} \langle j' |\nu(\vec{r}-\vec{r}') |j' \rangle = \notag\\
 \sum_{\vec{k}}\nu_{\vec{k}}.\text{exp}(i\vec{k}.\vec{r}).\sum_{j'=1}^{n} \int d^2 r' |\psi_{j'}(\vec{r}')|^2.\text{exp}(-i\vec{k}.\vec{r}').
\end{align}
Now in eq.~(2.16), the sum over $j'$ on the right-hand side is the expectation value of $(\Omega.\rho_{\vec{k}})$, eq.~(2.13), in the many-body state $\Psi$, eq.~(2.6). Thus, if we denote:
\begin{equation}
\langle \Psi| \rho_{\vec{k}} | \Psi \rangle = \langle \rho_{\vec{k}}\rangle,
\end{equation}
and the Hartree equation (2.7) and (2.8) becomes:
\begin{equation}
\left[ -\frac{1}{2\mu}\nabla^{2} + \Omega .\sum_{\vec{k}} \nu_{\vec{k}} . \langle \rho_{\vec{k}} \rangle.\text{exp}(-i\vec{k}.\vec{r}) \right]\psi_{j}(\vec{r}) = \varepsilon_{j}.\psi_{j}(\vec{r}).
\end{equation}

\subsection{\it graphene structure in a CDW}
\indent
\indent Graphene consists of a planar array of carbon atoms in a space filling set of open hexagons – see Fig.~1a. The crystal basis consists of two nearest neighbor atoms at sites (1) and (2), for example, in Fig.~1a, separated by the displacement vector $\vec{a}_{0,nn}$  of magnitude $a_0$ (not to be confused with the Bohr radius $a_0$  in eq.~(2.4)). The primitive translation vectors of the crystal lattice are $\vec{a}_1$ and $\vec{a}_2$, where:
\begin{align}
\vec{a}_1 = \frac{a}{2}\left(\sqrt{3}\hat{i}+\hat{j} \right), & \quad \vec{a}_2 = \frac{a}{2}\left(\sqrt{3}\hat{i}-\hat{j} \right)
\end{align}
and where:
\begin{equation}
a = |\vec{a}_1| = |\vec{a}_2|=\sqrt{3}.a_0,
\end{equation}
and $\hat{i}$ and $\hat{j}$ are unit vectors in the $x$ and $y$ directions. The reciprocal lattice basis vectors are:
\begin{align}
\vec{b}_1 = \frac{2\pi}{a}\left(\frac{1}{\sqrt{3}}\hat{i}+\hat{j} \right), &\quad \vec{b}_2 = \frac{2\pi}{a}\left(\frac{1}{\sqrt{3}}\hat{i}-\hat{j} \right),
\end{align}
with the defining property:
\begin{align}
(\vec{b}_i . \vec{a}_j) = 2\pi\delta_{ij}, &\quad b = |\vec{b}_j| = \frac{4\pi}{3a_0}.
\end{align}
The reciprocal lattice hexagonal unit is shown in Fig.~1b.  Notice the $(\pi/6)$ rotation between the hexagons in Figs.~1a and 1b.\\
\indent We now return to our Hartree equation~(2.18). For a CDW, we require that the scf consists of a wave pattern that has the geometrical shape of the atomic core pattern of graphene, Fig.~1a. We must keep in mind that our system, eq.~(2.18), contains no atomic cores. They have been subsumed into the effective electron mass parameter $\mu$, eq.~(2.2). It will now be the phonon mediated part of $\nu_{\vec{k}}$ in eq.~(2.18) that enables the system to have an scf in the form of a CDW. This may occur if the scf is dominated by $\vec{k}$ values that reflect the electronic structure of graphene. Thus consider:
\begin{equation}
\vec{k} = \pm \vec{K}_1,  \pm \vec{K}_2,  \pm \vec{K}_3,
\end{equation}
where:
\begin{equation}
\vec{K}_1 = \vec{b}_1,\quad \vec{K}_2 = \vec{b}_2,\quad \vec{K}_3 = (\vec{K}_1+\vec{K}_2)= (\vec{b}_1+\vec{b}_2).
\end{equation}
Vectors $\vec{K}_1$ and $\vec{K}_2$ ensure that our scf has the correct translational properties, and $\vec{K}_3$   is required to reflect the two-atom basis of graphene. All three vectors $\vec{K}_{j}$, $j=1,2,3$ have the same magnitude $K = b$, where then $K$ determines the scale $a_0$ of the wave pattern: see eq.~(2.22).\\
\indent The six wave vectors in eq. (2.23) taken together give three cosines. Consider:
\begin{equation}
F(\vec{r}) = \sum_{j=1}^3 \text{cos}(\vec{K}_j . \vec{r}).
\end{equation}
This wave pattern is shown as a contour map in Fig.~2. The large dark areas are centered on triangular mesh sites at wave pattern maxima. They tend toward hexagonal shape, representing each one’s relationship with six nearest neighbor minima. The latter show as light area, tending toward triangular shape, representing each one’s relationship with three nearest neighbor maxima. The minima correspond to atomic sites in graphene. The wave pattern therefore exactly reflects the potential seen by valence electrons in graphene.\\
\indent Now for our scf in eq.~(2.18) to behave like $F(\vec{r})$ in eq.~(2.25), we require the coefficients
\begin{equation}
\left( \nu_{\vec{k}}.\langle \rho_{\vec{k}}\rangle \right);\quad \vec{k} = \pm \vec{K}_j,\quad j=1,2,3
\end{equation}
all to be equal. Since $\nu(\vec{r})$ and $\rho(\vec{r})$ are real, we see that:
\begin{align}
\nu_{\vec{k}} = \nu_{-\vec{k}} \quad \text{and} \quad \rho_{\vec{k}} = \rho_{-\vec{k}}.
\end{align}
Furthermore, in Fig.~1b we see that vectors $\vec{K}_{j}$, $j=1,2,3$ are all equivalent in relation to the reciprocal lattice of the graphene structure, as well as in relation to each other. It follows that:
\begin{equation}
\left( \nu_{\vec{k}}.\langle \rho_{\vec{k}}\rangle \right) =\left( \nu_{K}.\langle \rho_{K}\rangle \right),\quad j=1,2,3
\end{equation}
The quantity $\nu_{K}$ is the second fundamental parameter of our model, the others being the effective band mass $\mu$, eq.~(2.2) and the CDW wave number $K$. This completes our model for a CDW having the graphene electronic structure, determined in terms of the Hartree equation (2.18), now in the form:
\begin{equation}
\left[ -\frac{1}{2\mu}\nabla^2 + 2\Omega^2.\nu_{K}.\langle \rho_{K} \rangle.\sum_{j'=1}^{3}\text{cos}(\vec{K}_{j'}.\vec{r}) \right]. \psi_{j}(\vec{r}) = \varepsilon_{j}.\psi_{j}(\vec{r}).
\end{equation}

\section{Method: Solving the Hartree Equation}

\subsection{\it tight-binding approximation}
\indent
\indent We now undertake the solution of eq.~(2.29). The discovery of the electronic structure of graphene dates from Wallace’s 1947 paper [2], before the era of electronic computation. Using only analytical methods applied in very simple approximations, Wallace was able to get the qualitative features of graphene correctly. Since that approach was successful, and fits with our objective of keeping our analysis as simple as possible, we shall follow Wallace’s work closely. To this end, we introduce the tight-binding approximation. It is briefly described by Wallace, and in more detail by Ashcroft and Mermin [6].\\
\indent We begin with Bloch functions of Wannier type. Because of the diatomic nature of our crystal basis, we view the crystal lattice as a superposition of two sublattices, A and B. Then the normalized Hartree eigenfunctions in eq.~(2.29) are:
\begin{equation}
\psi_{\vec{k}}(\vec{r}) = \frac{1}{\sqrt{n}}\lbrace \phi_{1,\vec{k}}(\vec{r}) \pm \phi_{2,\vec{k}}(\vec{r}) \rbrace
\end{equation}
where $\pm$ refer to states outside/inside BZI, and where:
\begin{align}
\phi_{1,\vec{k}}&= \sum_{A} \text{exp}(i\vec{k}.\vec{R}_A).X(\vec{r}-\vec{R}_A),\\
\phi_{2,\vec{k}} &=\sum_{B} \text{exp}(i\vec{k}.\vec{R}_B).X(\vec{r}-\vec{R}_B),
\end{align}
with $X(\vec{r})$ a normalized ground state atomic-like orbital. Note that we have now labelled $\psi_j$ by $\vec{k}$ values, where $\vec{k}$ ranges over the $n$ occupied Hartree eigenstates constituting the first Brillouin zone, BZI; similarly for $\varepsilon_{j}$ to become $\varepsilon_{\vec{k}}$. Now following Wallace, for the single particle energy, here $\varepsilon_{\vec{k}}$ and in Wallace $E$, we have:
\begin{equation}
\varepsilon_{\vec{k}} = \lbrace h'_{11}(\vec{k}) \pm |h'_{12}(\vec{k})|  \rbrace.
\end{equation}
In eq.~(3.4),
\begin{align}
h'_{11}(\vec{k}) &= \int d^2 r.\phi^{*}_{1,\vec{k}}(\vec{r}).h(\vec{r}).\phi_{1,\vec{k}}(\vec{r}),\\
h'_{12}(\vec{k}) &= \int d^2 r.\phi^{*}_{1,\vec{k}}(\vec{r}).h(\vec{r}).\phi_{2,\vec{k}}(\vec{r}).
\end{align}
\indent We now implement the tight-binding approximation by neglecting all but nearest-neighbor overlap integrals in sublattice A (or B) for the atomic-like orbitals $X(\vec{r})$ in $h'_{11}$, eq.~(3.5), and further, neglecting all but nearest-neighbor integrals between sublattices A and B in $h'_{12}$, eq.~(3.6).\\
\indent Still following Wallace, we introduce the periodic potential $V$ of the system. In the present case, this is the scf in eq.~(2.29):
\begin{equation}
V(\vec{r}) = 2\Omega^2. \nu_{K}. \langle \rho_K \rangle.\sum_{j=1}^3 \text{cos}(\vec{K}_j.\vec{r}).
\end{equation}
We also introduce the potential $U(\vec{r})$ for an isolated atomic-like orbital, determined from the minima of  $V(\vec{r})$. The precise analytical form of $U(\vec{r})$ will be given in the next subsection. It is the combination ($U$-$V$) that is central to the tight-binding approximation. Then from Wallace, and verifying his results, we obtain:
\begin{align}
h'_{11} &= \varepsilon_0 - 2\gamma'_0\left[ 2\text{ cos}\left(\frac{\sqrt{3}}{2}a k_x\right)\text{cos}\left(\frac{1}{2}a k_y \right) + \text{cos}\left(a k_y \right)\right] \\
|h'_{12}|&= \gamma_0 \left[1+4\text{ cos}^2\left(\frac{1}{2}a k_y \right)+4\text{ cos}\left(\frac{1}{2}a k_y \right)\text{cos}\left(\frac{\sqrt{3}}{2}a k_x \right) \right]^{1/2}
\end{align}
In eq.~(3.8):
\begin{align}
\varepsilon_0 &= \int d^2r.X^{*}(\vec{r}).h(\vec{r}).X(\vec{r}).\\
\gamma'_0 &= \int d^2r.X^{*}(\vec{r}).h(\vec{r}).X(\vec{r}-\vec{R}_A).
\end{align}
where $\vec{R}_A$ is any nearest-neighbor displacement of an atomic-like position from the origin in sublattice A. In eq.~(3.9):
\begin{equation}
\gamma_{0} = \int d^2r.X^{*}(\vec{r}).h(\vec{r}).X(\vec{r}-\vec{R}_B),
\end{equation}
where $\vec{R}_{B}$ has the same meaning in sublattice B. From Fig.~1a, we see that:
\begin{equation}
\vec{R}_A =\vec{a}_1,\quad \vec{R}_B = \vec{a}_0.
\end{equation}
From the tight-binding approximation we now have:
\begin{align}
\gamma'_0 = \int d^2r.X^{*}(\vec{r}).\left[U(\vec{r})-V(\vec{r}) \right].X(\vec{r}-\vec{a}_1),\\
\gamma_{0} = \int d^2r.X^{*}(\vec{r}).\left[U(\vec{r})-V(\vec{r}) \right].X(\vec{r}-\vec{a}_0).
\end{align}
Note that $\varepsilon_0$ is the energy of an electron in the atomic-like potential $U(\vec{r})$ . Also note that our definitions of $U$ and $V$ are vice versa to the definitions used by Ashcroft and Mermin.~[6] Eq.~(3.4), along with all the other formulae of this subsection, constitutes the tight-binding solution of our Hartree equation. However, we have yet to specify the atomic-like orbitals $X(\vec{r})$ in eqs.~(3.2) and (3.3).

\subsection{\it atomic-like orbitals}
\indent
\indent For our two-dimensional waveform $V(\vec{r})$, consisting of three harmonic waves, eq. (3.7), we approximate $V(\vec{r})$ in the region near one of its minima by the lowest order fit, namely a cylindrically symmetrical quadratic form, $U(\vec{r})$:
\begin{equation}
U(\vec{r}) = \left(-A+\frac{1}{2}B.r^2 \right).
\end{equation}
This potential will define our atomic-like orbitals $X(\vec{r})$. In eq.~(3.16) we have introduced a negative sign based on the assumption that $A$ is positive because the minima of $V(\vec{r})$ come from the minima of three cosines, all of which are negative at their minima. An electron in such a potential satisfies a two-dimensional cylindrical Schr\"{o}dinger equation:
\begin{equation}
\left[ -\frac{1}{2\mu}.\nabla^2 + \left(-A+\frac{1}{2}B r^2 \right) \right].X(\vec{r}) = \varepsilon'_0.X(\vec{r}).
\end{equation}
\indent
Consider a trial form for $X(\vec{r})$:
\begin{equation}
X(r) = \sqrt{\frac{2}{\pi}}.\alpha.\text{exp}(-\alpha^2 r^2),
\end{equation}
i.e. a normalized harmonic oscillator ground state wave function. If it is to be an eigenfunction in eq.~(3.17), we must have:
\begin{equation}
B = \frac{4\alpha^4}{\mu}
\end{equation}
and:
\begin{equation}
\varepsilon'_{0} = \left(\frac{2\alpha^2}{\mu} - A \right).
\end{equation}
\indent Now in eq.~(3.16), ($-A$) is the minimum value of our scf potential eq.~(3.7), and $B$ is its curvature. First consider (-$A$):
\begin{equation}
(-A) = 2\Omega.\nu_{K}. \langle \rho_{K}\rangle. \left[ \sum_{j=1}^{3} \text{cos}(\vec{K}_j \cdot \vec{r}) \right]_{\vec{r}=\vec{r}_1},
\end{equation}
where $\vec{r}_1$ is a minimum of $V(\vec{r})$. For example, atomic-like site (3) in Fig.~1a has position:
\begin{equation}
\vec{r}_1 = \frac{a}{2\sqrt{3}}\left(\hat{i}+\sqrt{3}\hat{j} \right).
\end{equation}
so:
\begin{equation}
A = 3.\Omega.\nu_{K}\langle \rho_{K}(\alpha)\rangle.
\end{equation}
It follows from eqs.~(3.20) and (3.23) that the atomic-like ground state energy is:
\begin{equation}
\varepsilon'_{0} = \left[ \frac{2\alpha^2}{\mu} - 3\Omega.\nu_{K}\langle \rho_{K}(\alpha)\rangle \right]
\end{equation}
\indent Now regarding $B$, the requirement that eq.~(3.17) should be an eigenvalue equation gives us eq.~(3.19). However, we also require $B$ to make $U(r)$, eq.~(3.16), fit the curvature of the potential $V(\vec{r})$ at its minimum. The second order terms from the expansion of $V(r)$ about $\vec{r}=\vec{r}_1$ are:
\begin{align}
\frac{1}{2}\sum_{\alpha,\beta=1}^{2}\left[ \frac{\partial^2}{\partial \chi_{\alpha}.\partial \chi_{\beta}}\left[ 2\Omega.\nu_{K}.\langle \rho_{K}\rangle.\sum_{j=1}^{3}\text{cos}(\vec{K}_{j}.\vec{r})\right].\chi_{\alpha}\chi_{\beta} \right]_{\vec{r}=\vec{r}_1} \notag\\
= \Omega.\nu_{K}.\langle \rho_{K}\rangle.\left( \frac{2\pi}{a} \right)^2 .r^2.
\end{align}
In eq.~(3.25), $\chi_{1}=x$, $\chi_2 =y$, $r^2=(x^2+y^2)$. Thus combining eqs.~(3.19) and (3.25) we have:
\begin{equation}
B = \frac{4\alpha^4}{\mu} = 2\Omega.\nu_{K}.\langle \rho_{K}\rangle.\left( \frac{2\pi}{a} \right)^2.
\end{equation}
This is a powerful constraint upon the solution of the Hartree equation, and as we shall see, it plays a central role in the analysis of the model system.

\section{The Self-Consistency Requirement}
\indent
\indent The equation (3.26) expresses the conditions that: (1) our assumed Bloch wave solution, eqs.~(3.1)-(3.3) and (3.18), is an eigenfunction of the Hartree equation, and (2) that our assumed atomic-like orbital $X(r)$ fits perfectly the minima of the scf potential. In that case, each Hartree eigenfunction depends on all $n$ of the eigenfunctions. For the ground state of the system, we take the $n$ Hartree eigenfunctions to be those of lowest single-particle eigenvalues $\varepsilon_{\vec{k}}$ : see eq.~(3.18). We therefore call eq.~(3.26) the self-consistency requirement.

\subsection{\it occupied $\vec{k}$-space region in the ground state}
\indent 
\indent In order to examine the scf more fully, we need explicit knowledge of $\langle \rho_{K} \rangle$. From eqs.~(2.13) and (2.6):
\begin{equation}
\langle \rho_{K}\rangle = \frac{1}{\Omega}.\sum_{j=1}^{n} \int d^2 r \ \psi^{*}_{j}(\vec{r}).\text{exp}(-i\vec{K}.\vec{r}).\psi_j(\vec{r}). 
\end{equation}
In Sec.~3.1, we have cast the Hartree eigenstates in terms of $\vec{k}$ values, eq.~(3.1) and later. Thus, consider the first Brillouin zone, denoted BZI: It consists of $\vec{k}$-space values within the mutual limits:
\begin{equation}
\vec{k}=\pm \frac{\vec{b}_i}{2},\quad i=1,2;\quad \frac{\vec{b}_3}{2} = \frac{\left(\vec{b}_1+\vec{b}_2 \right)}{2},\quad \text{or}
\end{equation}
\begin{equation}
\vec{k}= \pm \frac{\vec{K}}{2},\quad i=1,2,3,
\end{equation}
see eqs.~(2.23) and (2.24). This area, $A_{BZI}$, is therefore:
\begin{equation}
A_{BZI} = \frac{\sqrt{3}}{2}.b^2.
\end{equation}
The density $D(\vec{k})$ of $\vec{k}$-space points is:
\begin{equation}
D(\vec{k}) = \frac{\Omega}{(2\pi)^2}
\end{equation}
where $\Omega$ is the area of the Born-von Karmann region containing $n$ electrons in $N$ primitive unit cells of the Bravais lattice, each containing two electrons:
\begin{equation}
N = \frac{n}{2}
\end{equation}
Thus from eqs.~(4.4) and (4.5), the number of one-electron states in BZI including spin is:
\begin{equation}
2.A_{BZI}.D(\vec{k}) = 2.\left(\frac{\sqrt{3}}{2}.b^2 \right).\left( \frac{\Omega}{(2\pi)^2} \right).
\end{equation}
Now $\Omega$ is:
\begin{equation}
\Omega = N.|\vec{a}_1 \times \vec{a}_2| = n a^2 . \frac{\sqrt{3}}{4},
\end{equation}
from eqs.~(4.6) and (2.19). We obtain the value of $b$ from eq.~(2.21), giving for eq.~(4.7):
\begin{equation}
2.A_{BZI}.D(\vec{k}) = n.
\end{equation}
This verifies that the number of electrons in our CDW exactly fills BZI. Wallace [1] has shown that all the Hartree eigenstates in BZI have lower energy than those outside.

\subsection{\it the role of average electronic density in CDW}
The average electronic density in  $\vec{r}$-space, $\rho_0$, is:
\begin{equation}
\rho_0 = \frac{n}{\Omega}.
\end{equation}
From eq.~(4.8), and the relationship between $a$ and $K$, eqs.~(2.19) and (2.21), we find:
\begin{equation}
\rho_0 = \sqrt{3} \left( \frac{K}{2\pi}\right)^2,
\end{equation}
or:
\begin{equation}
K = (2\pi).\left( \frac{\rho_0}{\sqrt{3}} \right)^{1/2}.
\end{equation}
In words, eq.~(4.11) tells us that if a CDW is required with a given value of $K$, then the average density must have the value specified there. On the other hand, if our system has a given density $\rho_0$, the only $K$-value which can form in a CDW is given by eq.~(4.12).
\subsection{\it evaluation of $\langle \rho_{K}\rangle$ }
We must now evaluate $\langle \rho_{K} \rangle$, eq.~(4.1). There, we convert the sum over $j$ to an integral over  $\vec{k}$-space in BZI, ($\times 2$) for spin:
\begin{equation}
\sum_{j=1}^{n} \rightarrow 2.\int_{BZI} d^2 k. D(\vec{k}).
\end{equation}
In eq.~(4.1), expressed in terms of $\vec{k}$, $\psi_{\vec{k}}(\vec{r})$ is given by eqs.~(3.1)-(3.3) and (3.18), with the sums over $\vec{R}_A$ and $\vec{R}_B$ limited according to our application of the tight-binding approximation. The resultant evaluation has been determined analytically by using MATLAB and MAPLE independently, with the result:
\begin{equation}
\langle \rho_{K}\rangle = \frac{1}{a^2}.f(w),
\end{equation}
where:
\begin{equation}
w = (\alpha a)^2,
\end{equation}
and
\begin{equation}
f(w) = \text{exp}\left(-\frac{2\pi^2}{3w} \right).\left[\frac{1}{2\pi^2}\left(6\sqrt{3}+4\pi \right).\text{exp}\left(-\frac{w}{6}\right)-\frac{2}{\sqrt{3}} \right].
\end{equation}

\subsection{\it self-consistency}
\indent
\indent We now evaluate the scf condition expressed by eq.~(3.26), using the notation $w$, eq.~(4.15), the expression for $\langle \rho_{K} \rangle$ , eqs.~(4.14)-(4.16), and:
\begin{equation}
\nu_K = \frac{1}{\Omega} \int d^2 r.\nu(r).\text{exp}(-i\vec{K}.\vec{r}).
\end{equation}
The result is:
\begin{equation}
w^2 = (\mu.\nuktilde).f(w),
\end{equation}
where:
\begin{equation}
\nuktilde = 2\pi^2.\int d^2 r.\nu(r).e^{i\vec{K}.\vec{r}} = 2\pi^2.\Omega.\nu_{K}.
\end{equation}
Eq.~(4.18) is the self-consistency condition required of a solution to the Hartree equation within the context of the tight-binding approximation as applied to a CDW. Only values of $w$ that satisfy eq.~(4.18) may be used in calculating properties of the CDW system, such as the total energy.\\
\indent From its definition, eq.~(4.15), $w$ must be positive. From eq.~(4.18), several possibilities arise. Basically, it is required that:
\begin{equation}
(\mu.\nuktilde).f(w)>0.
\end{equation}
We shall exemplify the possible solutions by considering here only the case:
\begin{equation}
\mu >0 .
\end{equation}
We remark that the case $\mu < 0$ means that the electron gas consists of hole-like, rather than electron-like, quasiparticles.\\
\indent Now from eqs.~(4.20) and (4.21), two possibilities remain:
\begin{align}
\text{Case (i):} \quad & \nuktilde>0 \text{ and } f(w)>0, \\
\text{Case (ii):} \quad & \nuktilde<0 \text{ and } f(w)<0.
\end{align}
Consider Case (i). From eq.~(4.16) we find that $f(w)$ is positive only for:
\begin{equation}
0 \leq w \leq 0.0435.
\end{equation}
For $w$ = 0.0435, $\alpha \approx (0.2/a) \approx 1/(5a)$. Now $\alpha$ is a measure of the range $R$ of the Gaussian atomic-like orbital, namely the distance at which the orbital’s amplitude is $e^{-1} \approx 0.37$  times its maximum value. Thus we find that for $f(w)>0$:
\begin{equation}
w = (\alpha a)^2 = (a/R)^2 < 0.0435,
\end{equation}
whence:
\begin{equation}
R > 8.3 a_0,
\end{equation}
where $a_0$ is the nearest-neighbor distance in the graphene structure. However, implicit in the tight-binding approximation is the assumption that the atomic-like orbitals are well-localized within a range $R \lesssim a_0$. Thus the solution with $f(w)>0$ is inconsistent with the present model. We conclude that Case (i) is unphysical. At this point we therefore conclude, perhaps not surprisingly, that with $\mu>0$, $\nuktilde$ and therefore   $\nu_{K}$ must be negative, i.e. attractive. It means that, at wave number $K$, the phonon-mediated part of the pairwise interaction must be attractive, and must overwhelm the corresponding component of the Coulomb repulsion so that the total interaction is also attractive.\\
\indent Now consider Case (ii), with $\nuktilde <0$ and $f(w)<0$. The latter condition is found to be valid for $w >$ 0.0435. Fig.~3 shows the plot of $f(w)$ vs. $w$, with a horizontal asymptote at $|f(w)| =$ 1.1547. We now express the scf condition in terms of $w$ as a function of $x=(\mu.|\nuktilde|)$, with $\mu>0$:
\begin{equation}
\frac{w^2}{|f(w)|} = x. 
\end{equation}
This is plotted in Fig.~4, where the coordinates for a minimum value of $x$ are given, showing that self-consistency is not possible for $x \lesssim $ 143.5. Fig.~4 will be further interpreted in Sec.~5.
\section{Total Energy of the CDW System}
\indent
\indent The total energy $E_n$ of our $n$-electron system has been introduced in terms of the Hartree approximation: see eq.~(2.10). We have chosen to solve the Hartree equation by following Wallace’s approach, namely the tight-binding approximation, Sec.~3.1. In eq.~(2.10), the first term is converted to:
\begin{equation}
\sum_{j=1}^{n} \varepsilon_{j} = 2 \int_{BZI} d^2k.\varepsilon(k).D(\vec{k}).
\end{equation}
The second term in eq.~(2.10), correcting for double counting of pairwise interactions in $\varepsilon_{j}$, is evaluated from eq.~(2.9). Now, the pairwise interaction energy $V_2$ of our system is:
\begin{equation}
V_2 =\frac{1}{2} \sum_{j,j'=1}^{n} \langle \Psi | \nu(\vec{r}_{j}-\vec{r}_{j'}) | \Psi \rangle.
\end{equation}
Half of this must be subtracted from the sum over $\varepsilon_j$.\\
\indent Let us Fourier analyse $\nu(\vec{r})$ in the union of spin space and configuration space:
\begin{equation}
\nu(\vec{r}_{j}-\vec{r}_{j'}) = 2 \sum_{\vec{k}} \nu_{\vec{k}}.\text{exp}\left[i\vec{k}.\left(\vec{r}_j-\vec{r}_{j'} \right) \right]
\end{equation}
Then:
\begin{equation}
V_2 = \frac{1}{2}.2.\sum_{\vec{k}}\nu_{\vec{k}}.\left|\sum_{j}\int d^2 \vec{r}_j.\text{exp}(i\vec{k}.\vec{r}_j).|\psi_{j}(r_j)|^2 \right|^2.
\end{equation}
As explained in Sec.~2.3, the integral in eq.~(5.4) is simply $\left(\Omega.\langle \rho_{\vec{k}} \rangle \right)$. Our correction term is therefore:
\begin{equation}
\left( -\frac{1}{2}.V_2 \right) = -\frac{1}{2} \sum_{\vec{k}(BZI)} \Omega^2.|\langle \rho_{\vec{k}} \rangle|^2.\nu_{\vec{k}}.
\end{equation}
For our CDW case, the values of $\vec{k}$ are limited to $\pm \vec{K}_{j}$, $j = 1,2,3,$ as in eqs.~(2.23) and (2.24) of Sec.~2.5, so now:
\begin{equation}
\left( -\frac{1}{2}.V_2 \right) = -3\Omega^2 |\langle \rho_{K} \rangle|^2.\nu_{K}.
\end{equation}
The total energy expression $E_{n}$, eq.~(2.10) is therefore:
\begin{equation}
E_{n} = \left[ 2\int_{BZI} d^2 k.\varepsilon(\vec{k}).D(\vec{k})-3\nu_K \Omega^2.|\langle \rho_{K} \rangle|^2 \right].
\end{equation}
The first term, eq.~(5.1), is made explicit by introducing the results of Sec.~3. The result can be expressed analytically, with the exception of one integral, of the form:
\begin{equation}
I(a) \sim \int_{BZI} d^2 k  \left[1+4\text{ cos}^2\left(\frac{1}{2}a k_y \right)+4\text{ cos}\left(\frac{1}{2}a k_y \right)\text{cos}\left(\frac{\sqrt{3}}{2}a k_x \right) \right]^{1/2}
\end{equation}
By numerical integration, a constant $c_0$ is introduced, where:
\begin{equation}
c_0 = 17.94479903.
\end{equation}
The final result is:
\begin{align}
E_{n} = \frac{4\Omega}{\sqrt{3}.a^2}.\left( \frac{2\alpha^2}{\mu} -3\Omega.\frac{\nu_{K}}{2}.\langle \rho_{K} \rangle \right)+c_0 .\Omega^2.\nu_{K}.\langle\rho_{K} \rangle.\text{exp}\left(-\frac{(\alpha a)^2}{6}\right).\notag\\
\left[ \frac{6}{(\pi a)^2}-\frac{1}{a^2}\left(\frac{1}{3}+\frac{2}{(\alpha a)^2} \right)+\frac{2}{(\pi a)^2}.\text{exp}\left(-\frac{2\pi^2}{3(\alpha a)^2}\right) \right] - 3\Omega^2.\nu_{K}.|\langle \rho_{K} \rangle|^2.
\end{align}
We now can express $E_n$ in terms of the material parameters $\mu$, $\nu_K$ (or $\tilde{\nu}_{K}$ ) and $\rho_0$, which determines $K$, eq.~(4.11) or (4.12) in a canonical form by introducing a set of previously defined relationships, as follows. For $\Omega$, we use eq.~(4.8). For $\alpha$, we use eq.~(4.15). For $\langle \rho_{K} \rangle$, we use eq.~(4.14) with (4.16). For $\nu_{K}$, we use eq. (4.19) to introduce $\tilde{\nu}_{K}$.
\begin{equation}
\nuktilde = 2\pi \Omega^2.\nu_K.
\end{equation}
For $a$ we use eqs.~(2.24) and (2.21) to get:
\begin{equation}
\frac{1}{a^2} = \frac{3}{4}\left(\frac{K}{2\pi} \right)^2.
\end{equation}
With these substitutions we get, from eq.~(5.10), the total energy $E_n$:
\begin{equation}
E_{n} = n \left(\frac{K^2}{2\mu}\right) . \left(\frac{3}{8\pi^2}\right) . G(w),
\end{equation}
where:
\begin{align}
G(w) = 2w +\frac{\sqrt{3}}{4\pi^2}.\left(\mu.|\nuktilde| \right).|f(w)|.\notag\\
\left[-4.\sqrt{3}+c_0\left[\frac{6}{\pi^2}-\left(\frac{1}{3}+\frac{2}{w} \right)+\frac{2}{\pi^2}\text{exp}\left(-\frac{2\pi^2}{3w}\right) \right].\text{exp}\left(-\frac{w}{6}\right) +3|f(w)| \right].
\end{align}
This expression for $G(w)$ may be evaluated subject to the scf condition given in eq.~(4.27), in terms of the quantity $x$ introduced in Sec.~4:
\begin{equation}
x = (\mu.|\nuktilde|),\text{ } \mu>0.
\end{equation}
This is done by using $w$ as a function of $x$, as plotted in Fig.~4, resulting in $G(w)$ becoming $\Gamma(x)$, shown in Fig.~5.
\begin{equation}
E_{n} = n\left( \frac{K^2}{2\mu} \right) \left( \frac{3}{8\pi^2} \right).\Gamma(x).
\end{equation}
We can see in eq.~(5.16) that $E_n$ is negative whenever $\Gamma(x)$ is negative, if $\mu$ is positive. From Fig. 5 we can see that $\Gamma(x)$, and therefore $E_n$, is negative for $x > $189.\\
\indent We shall now show that the limitations of the tight-binding approximation translate into a limitation on the range of $x=\mu.|\nuktilde|$ within which our results may be expected to be valid. Roughly speaking, tight-binding implies that the range $R=\alpha^{-1}$ of the atomic-like orbitals (see eq.~(3.18), and discussion following eq.~(4.24)) should be significantly less than the nearest neighbor distance $a_0$ between pairs of such orbitals. In turn, the scf condition, eq.~(4.27) with the definition $w=(a \alpha)^2 = (a/R)^2$, gives a direct relationship between $R$ and $x$.\\
\indent To illustrate the situation, we consider:
\begin{equation}
0.25 < \frac{R}{a_0} < 0.5.
\end{equation}
The case $R=0.5 a_0$ is a liberally large value, where the atomic-like orbital has an amplitude $A$ halfway between nearest-neighbor sites of  $\sim 0.37 A_0$, where $A_0$ is the atomic orbital maximum. For the case of   $R=0.25 a_0$, the ratio $(A/A_0)$ is $\sim$0.02, i.e. the overlap of two orbitals might be considered to be negligible. The criterion expressed in eq.~(5.17) can be given in terms of $x$ by using the scf condition, to read:
\begin{equation}
250 \leq x \leq 2,290.
\end{equation}
These limits, relating to the tight-binding approximation, are shown in Fig.~5 by vertical dashed lines.
\section{The Spatially Uniform State}
\indent
\indent We recall that from eq.~(2.3) our model consists only of electrons with kinetic energy and with pairwise interactions that have both Coulomb repulsion and phonon-mediated attraction. We shall now show that there is a simple solution to this model, apart from the CDW solution developed heretofore. We shall examine the kinetic energy part first, and then see what the interactions contribute to the total energy.\\
\indent Eigenstates of the single-particle kinetic energy operator are:
\begin{equation}
\psi_{j}(\vec{r}) = \Omega^{-1/2}.\text{exp}(i\vec{k}_j \cdot \vec{r}); \quad \varepsilon_{j} = \frac{k^{2}_{j}}{2\mu};
\end{equation}
see eqs.~(2.7)-(2.8). It is obvious that the spatial distribution of electrons in this $n$-particle system is uniform, i.e. independent of particle positions or spins. The ground state of this non-interacting system consists of vectors $\vec{k}_j$ that fill the Fermi circle $k=k_{F}$, where $k_F$ is defined by:
\begin{equation}
2\int\limits_{k=0}^{k_F} d^2k.D(\tilde{k}) = n;
\end{equation}
see eq. (4.13). From this we find:
\begin{equation}
k_{F} = \left(2\pi.\rho_0 \right)^{1/2}
\end{equation}
Thus the total ground state kinetic energy $E^{(0)}_n$ is:
\begin{equation}
E^{(0)}_{n} = 2\int\limits_{k=0}^{k_F} d^2k.D(\vec{k}) . \frac{k^2}{(2\mu)} = n\left(\frac{\pi. \rho_0}{2\mu} \right).
\end{equation}
\indent
Now consider the pairwise interactions' contribution to the uniform density state. Up to now, and throughout, as in eq.~(5.2), we have included terms $j=j'$ so that in the Hartree equation each electron sees the same scf, to conform to the principle that identical particles must be indistinguishable. We acknowledge, however, that this principle is already violated in the Hartree approximation by the assumed form of the $n$-particle wave function, eq.~(2.6), where each particle exists exclusively in a single basis function state, i.e. the $j$th particle is represented by $\psi_{j}$, and is therefore distinguishable from particle $j' \neq j$. In the uniform state, with:
\begin{equation}
\Psi^{(0)}_n = \prod_{j=1}^{n} \Omega^{-1/2}.\text{exp}(i\vec{k}_j \cdot \vec{r}_j),
\end{equation}
the pairwise interaction energy $V^{0}$ is:
\begin{equation}
V^{(0)} = \langle \Psi_{n}^{(0)} | \frac{1}{2} \sum_{j,j'=1} \nu(\vec{r}_{j}-\vec{r}_{j'})| \Psi_{n}^{(0)} \rangle.
\end{equation}
The contribution to $V^{(0)}$  from terms with $j=j'$ is simply:
\begin{equation}
\sum_{j=1}^{n} \nu(0) = n.\nu(0).
\end{equation}
\indent
Consider now a single term with $j \neq j'$. For $\nu(\vec{r}_{j}-\vec{r}_{j'})$, write the Fourier series:
\begin{equation}
\nu(\vec{r}_{j}-\vec{r}_{j'}) = \sum_{\vec{k}} \nu_{\vec{k}}.\text{exp}(i\vec{k}.\vec{r}_j).\text{exp}(-i\vec{k}.\vec{r}_{j'}).
\end{equation}
\noindent
Such a term contributes to $V^{(0)}$:
\begin{equation}
\nu_{\vec{k}}.\langle \psi_{j}|\text{exp}(i\vec{k}.\vec{r}_j)|\psi_{j} \rangle.\langle \psi_{j'}|\text{exp}(-i\vec{k}.\vec{r}_{j'})|\psi_{j'}\rangle.
\end{equation}
\noindent
Since here $|\psi_j|^2=1$, and since in our model $\nu_{\vec{k}}$  is limited to $\vec{k}$-values given in eqs.~(2.23) and (2.24), the relevant terms from eq.~(6.9) are symmetrical in pairs, constituting $\text{cos}(\vec{k}_j.\vec{r}_j)$ type of terms. Then in eq.~(6.9), we are left with terms like:
\begin{equation}
\langle \psi_{j}|\text{cos}(\vec{k}.\vec{r}_j)|\psi_{j} \rangle = 0.
\end{equation}
\noindent
It follows that eq.~(6.7) is the total value of $V^{(0)}$:
\begin{equation}
V^{(0)} = n.\nu(0).
\end{equation}
\noindent
Combining this with the non-interacting particle energy, eq.~(6.4), we have
\begin{equation}
E^{(0)}_{n} = n\left( \frac{\pi \rho_0}{2\mu}+\nu(0) \right).
\end{equation}
\noindent
This is not useful, however, because $\nu(0)$ is related to the self energy of an electron, which can be given meaning only in the context of the quantum field picture of the particle, which is beyond the scope of our very elementary model. We can only assume that the second term in eq.~(6.12) is negligible compared to the first, so that $E^{(0)}_n$ as given in eq.~(6.4) is the total energy. In general, the limitations of Hartree-based methods that we have mentioned, namely distinguishability of identical particles, and electron self-energy, can be overcome easily by using the Hartree-Fock approximation (HF), in which neither problem exists. In HF, the exchange energy is also included, rigorously, and correlation can also be included approximately if it is merely perturbative. While HF is much more work to implement, it also represents a large step in rigor compared to the Hartree approximation. It might therefore be considered for follow-up to the present work.

\section{The Stability Condition}
\indent
\indent For the CDW state of the system to be the stable ground state, its energy must be the lowest of all states. In particular, it must have lower energy than $E^{(0)}_{n}$, the energy of the uniform density state, eq.~(6.4). In Sec.~5 and in Fig.~5 we have seen that $E_n$ is negative when $\Gamma(x)$ is negative, and $\Gamma(x)<0$  for $x > $189. We have also shown that our tight-binding results are probably not valid for $x$-values that are much less than $x$ = 250. Thus, within the range of validity of our model, $E_n$ is generally negative. At the same time, $E^{(0)}_n$, eq.~(6.4), is always positive, for $\mu>0$. Thus the requirement, that $E_n < E^{(0)}_n$ for stability of the CDW state relative to the uniform state, is generally satisfied for our model.\\
\indent We can relate $E_n$ to $E^{(0)}_n$ by using the relationship between $K$ and $\rho_0$, eq.~(4.12). Then we have:
\begin{equation}
E_n = E_n^{(0)}. \frac{\sqrt{3}}{2\pi}.\Gamma(x).
\end{equation}
This enables us to introduce the fractional deviation $\delta E_n$ of $E_n$ from $E^{(0)}_n$:
\begin{equation}
\delta E_{n} = \frac{(E^{(0)}_n-E_n)}{E^{(0)}_{n}} = \left[1 - \frac{\sqrt{3}}{(2\pi)}.\Gamma(x) \right].
\end{equation}
The stability requirement, obtained in the previous paragraph, that  $E^{(0)}_n > E_n$, now becomes $\delta E_n >0$, which from eq.~(7.2) comes down to:
\begin{equation}
\Gamma(x) < \frac{(2\pi)}{\sqrt{3}} \approx 3.628.
\end{equation}
This is the basic analytical condition for stability of the CDW state relative to the uniform-density state. The upper limit of 3.6 on $\Gamma(x)$ is manifestly positive. However, in Fig.~5 we see that $\Gamma(x)$ is negative for $x >$ 189, and in particular it is negative throughout the range of $x$-values within which our tight-binding method may be expected to be valid. This range has a lower limit of $x$ = 250 at least, as expressed in eq.~(5.18), and as seen in Fig.~5. We therefore see that the stability condition for our model system is satisfied conclusively if only the scf condition is satisfied.

\section{Summary and Conclusions}
\subsection{\it summary}
\indent
\indent We have raised the question whether the scf for the electrons in a two-dimensional CDW can mimic the potential due to the atomic cores in graphene. If so, then the band structure in the CDW state will be qualitatively the same as it is in graphene, and one might expect it to have at least some of the properties of graphene. The mathematical and physical requirements for such a CDW system may define a class of materials that are quite different from graphene, while possessing some of graphene’s attractive properties.\\
\indent We have adopted a simple model for a CDW in which the hamiltonian for an $n$-electron system consists of kinetic energy and pairwise electron interactions (pwi). The kinetic energy includes the periodic potential due to the atomic cores of our host crystal by including an effective band mass, denoted $\mu$. The pwi collectively define a self-consistent field which, for a model CDW, must be periodic. The periodicity will have minima and maxima that correspond to the atomic-core pattern and its interstices in graphene. The scf introduces, in addition to $\mu$, the physical parameters $\tilde{\nu}_{K}$ and $K$, representing respectively the amplitude and wave number of the CDW.\\
\indent The wave number $K$ is determined by the average electronic density $\rho_0$. We solve this many-electron problem by using very simple approximations, based as much as possible on Wallace’s original method. In this way, we determine the scf condition for the existence of an appropriate CDW. We develop an algorithm for an $n$-electron system, leading to an analytical formula for the total energy. Within the same model, we evaluate the total energy of a uniform-density state. We then determine in detail the criteria for the stability of the CDW relative to the uniform density state.

\subsection{\it conclusions}
The existence of a two-dimensional CDW mimicking graphene electronic structure relies on the CDW having six dominant terms all of the same amplitude $\tilde{\nu}_K$ in its Fourier series, that reflect the internal symmetry of graphene, eq.~(2.23) and (2.24). In general, this must be accomplished by experimental manipulation of basic physical parameters $\mu$, $\tilde{\nu}_K$ and $K$ (or $\rho_0$), qualitatively as has been done to induce superconductivity and CDW in graphene.\\
\indent We find that, in our model, a self-consistent solution requires that $\tilde{\nu}_K$ should be negative, indicating that at wavenumber $K$, the pairwise interaction, consisting of both Coulomb repulsion and phonon-induced attraction, must be attractive in total. Explicitly, however, we find that $|\tilde{\nu}_K|$ must be larger than a specific finite value: see Fig.~4. We find that the CDW has lower total energy than the uniform-density state, and is therefore the stable ground state (among the two of them). We have studied the lower-bound limitations placed on $|\tilde{\nu}_K|$ for fixed values of $\mu$ and $\rho_0$ by the scf, the stability condition, and the tight-binding approximation, successively, and find that there is no contradiction among them. The ultimate limitation is that imposed by the tight-binding approximation.\\
\indent The results of this study of a tight-binding model of a CDW having the graphene electronic structure support the idea of developing such a class of materials. On the theoretical side, a similar study of a weak-binding model appears feasible and might be useful.

\section*{Acknowledgments}
\indent
\indent Author J. M. V. gratefully acknowledges, in his capacity as Senior Scholar, the support of the Faculty of Science and of the Department of Physics and Astronomy at the University of Manitoba, without which this work could not have been done. Author O. J. H. was supported in part by the Canadian Natural Sciences and Engineering Council (NSERC). We thankfully acknowledge personal support for this work by D. S. Xue of Lanzhou University and by R. Pandey of Michigan Technological University.

\newpage
\section*{Figure Captions}
Fig. 1a.	Hexagonal unit of graphene crystal, with basis sites (1) and (2) and separation $\vec{a}_0$, primitive translation vectors $\vec{a}_{1}$ and $\vec{a}_2$, and $x$-$y$ coordinate axes.\\

\noindent
Fig. 1b.	Hexagonal first Brillouin zone in the reciprocal lattice of the graphene crystal, with primitive translation vectors $\vec{b}_1$ and $\vec{b}_{2}$, and $k_x$-$k_y$ coordinates.\\

\noindent
Fig. 2.	Wave pattern, eq.~(2.25), of our two-dimensional CDW model with the graphene structure.\\

\noindent
Fig. 3.	$f(w)$, eq.~(4.16), plotted vs. $w$.\\

\noindent
Fig. 4.	Plot of $w$ vs. $x$, showing the trends of range $R$, for the case $\mu=1$. See eq.~(4.27).\\

\noindent
Fig. 5.	Plot of $\Gamma(x)$ vs. $x=(\mu |\tilde{\nu}_{K}|)$ for the case $\mu=1$, in the formula for total CDW energy $E_n$, eqs.~(5.14), (5.15) and (5.16): see discussion following eq.~(5.16). Shown by vertical dashed lines are plausible limits on $R$ for validity of the tight-binding approximation, eqs.~(5.17) and (5.18).

\newpage
\begin{figure}[h]
\centering
\includegraphics[scale=0.35]{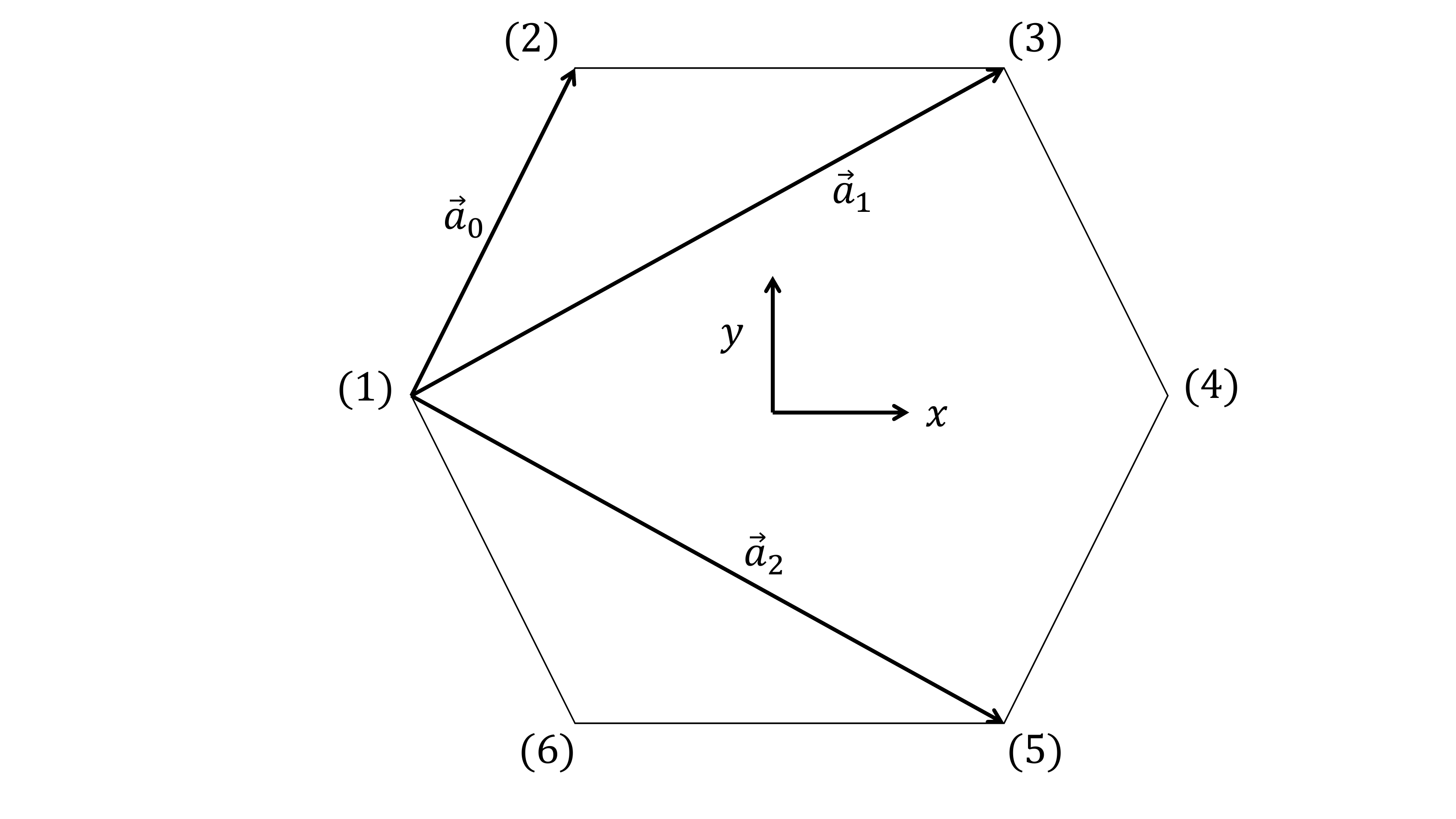}
{\caption*{{\bf Fig. 1a}}}
\end{figure}

\begin{figure}[h]
\centering
\includegraphics[scale=0.35]{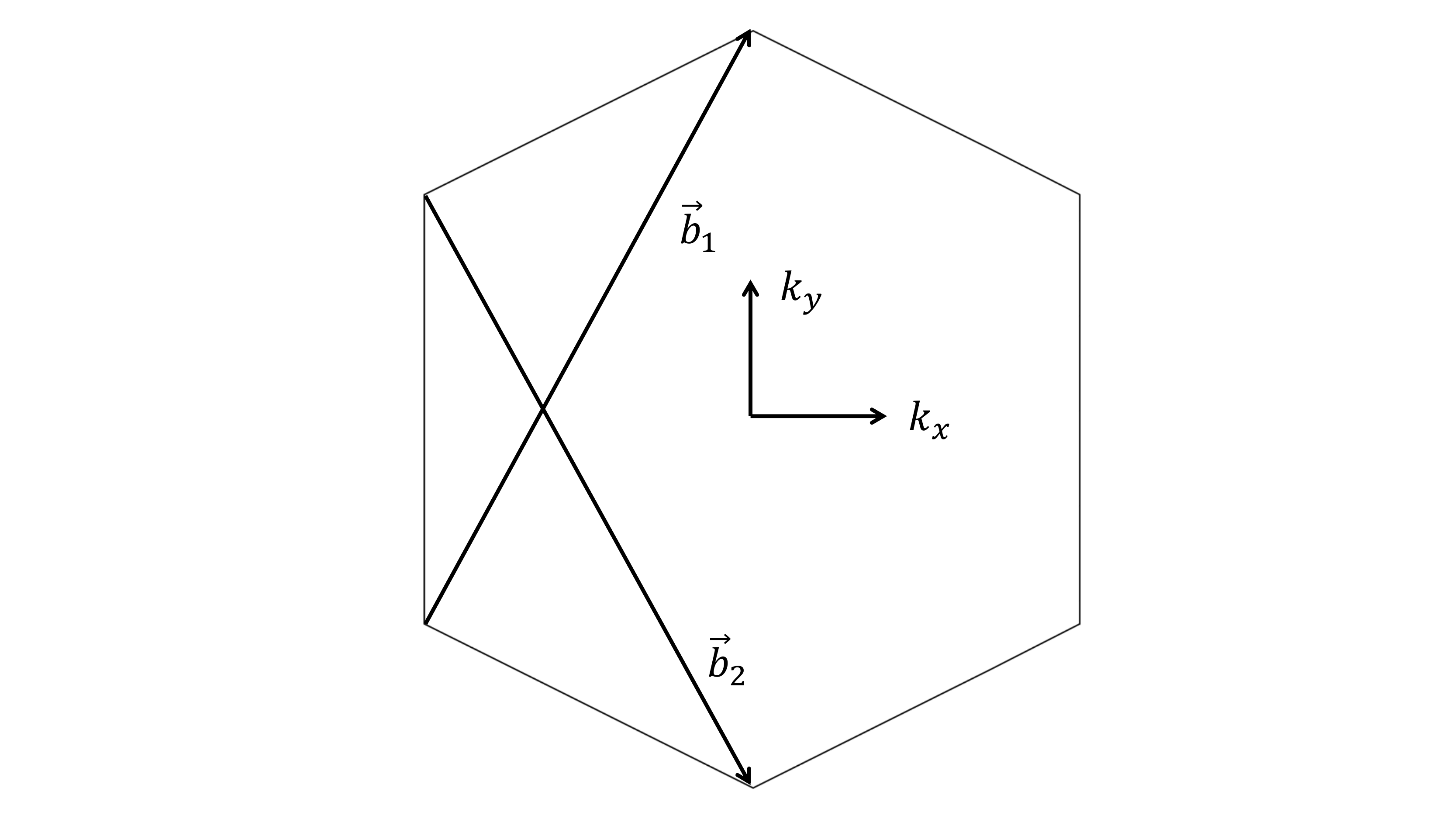}
{\caption*{{\bf Fig. 1b}}}
\end{figure}

\newpage
\begin{figure}[h]
\centering
\includegraphics[scale=0.7]{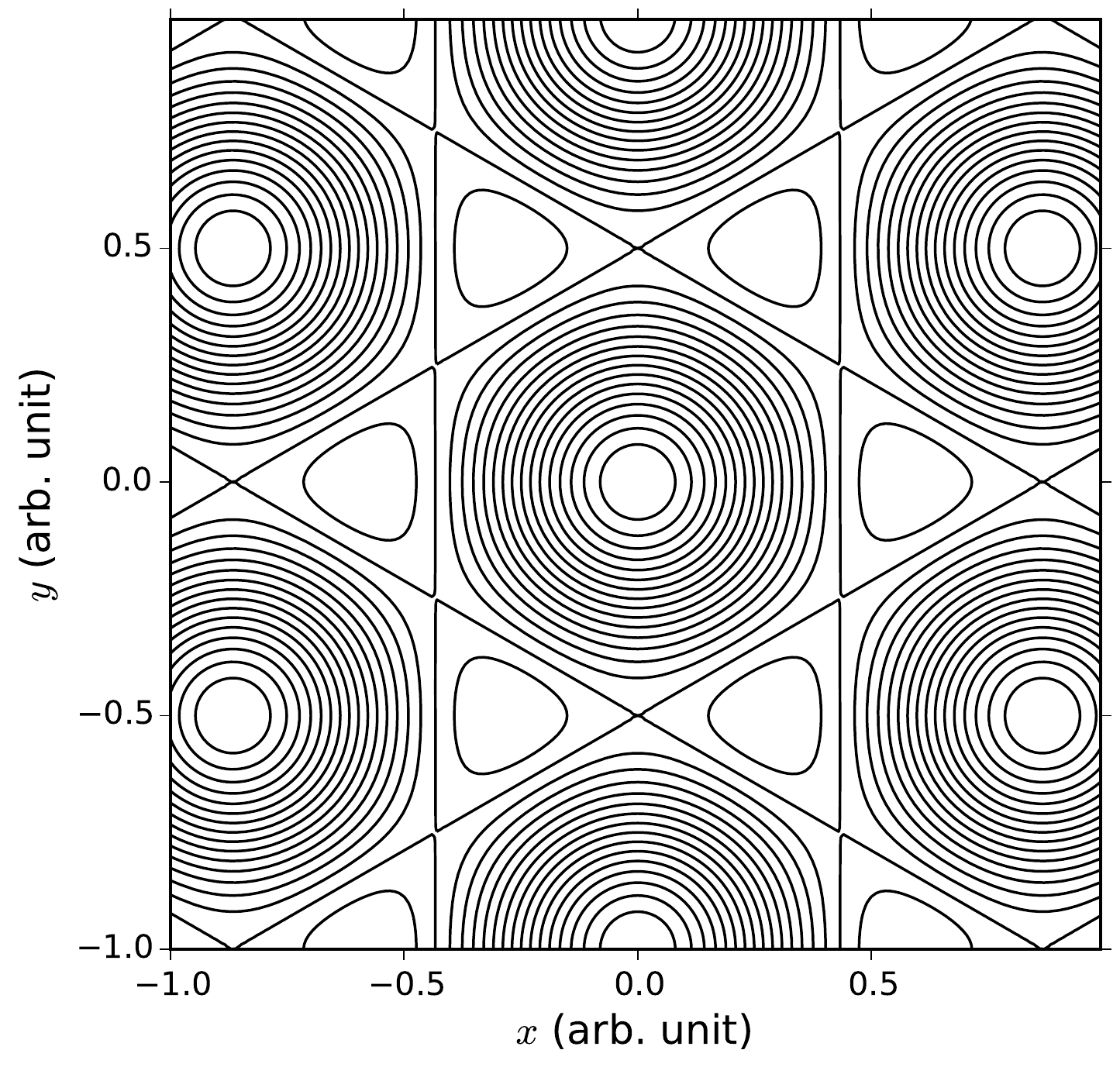}
{\caption*{{\bf Fig. 2}}}
\end{figure}

\newpage
\begin{figure}[h]
\centering
\includegraphics[scale=0.7]{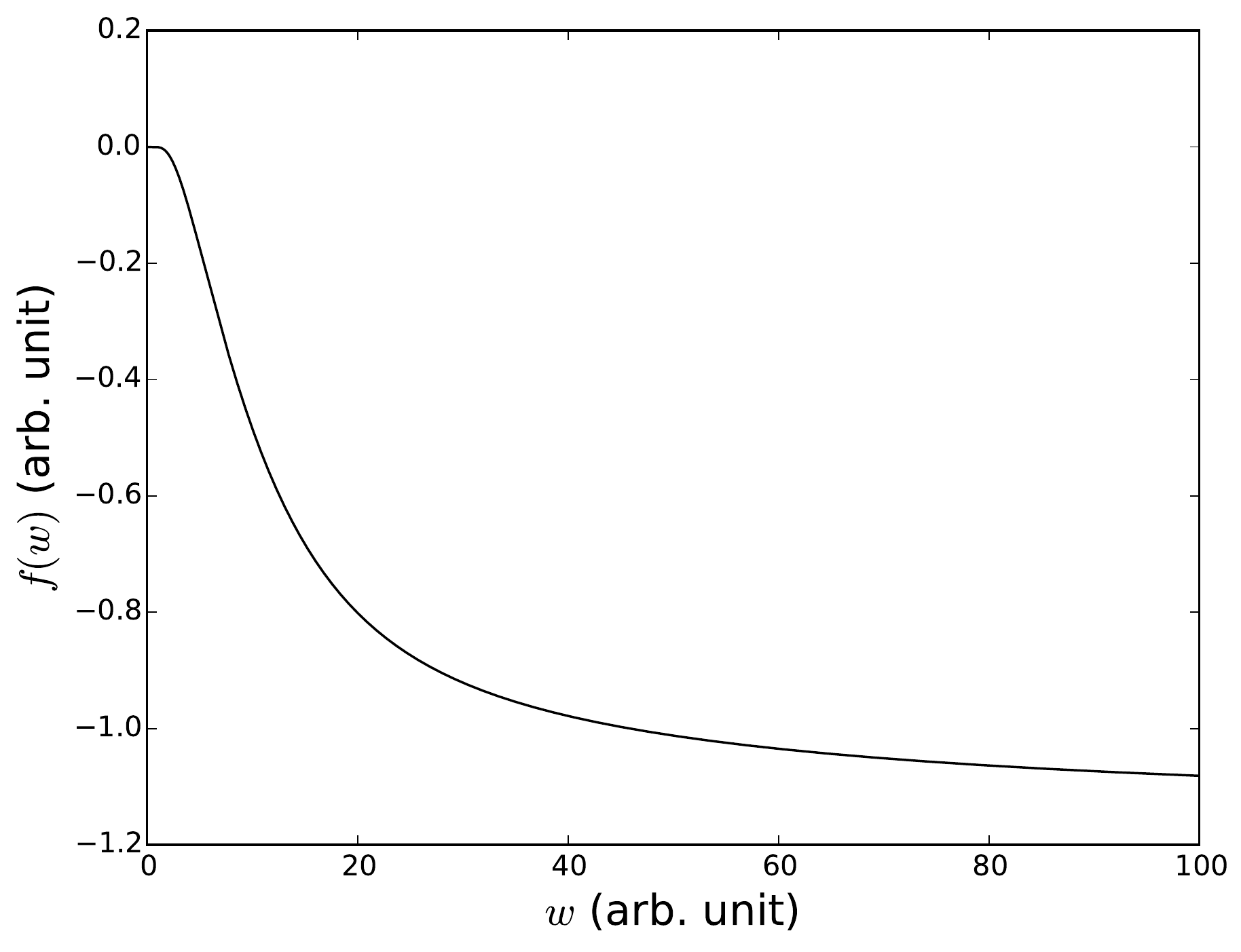}
{\caption*{{\bf Fig. 3}}}
\end{figure}

\newpage
\begin{figure}[h]
\centering
\includegraphics[scale=0.7]{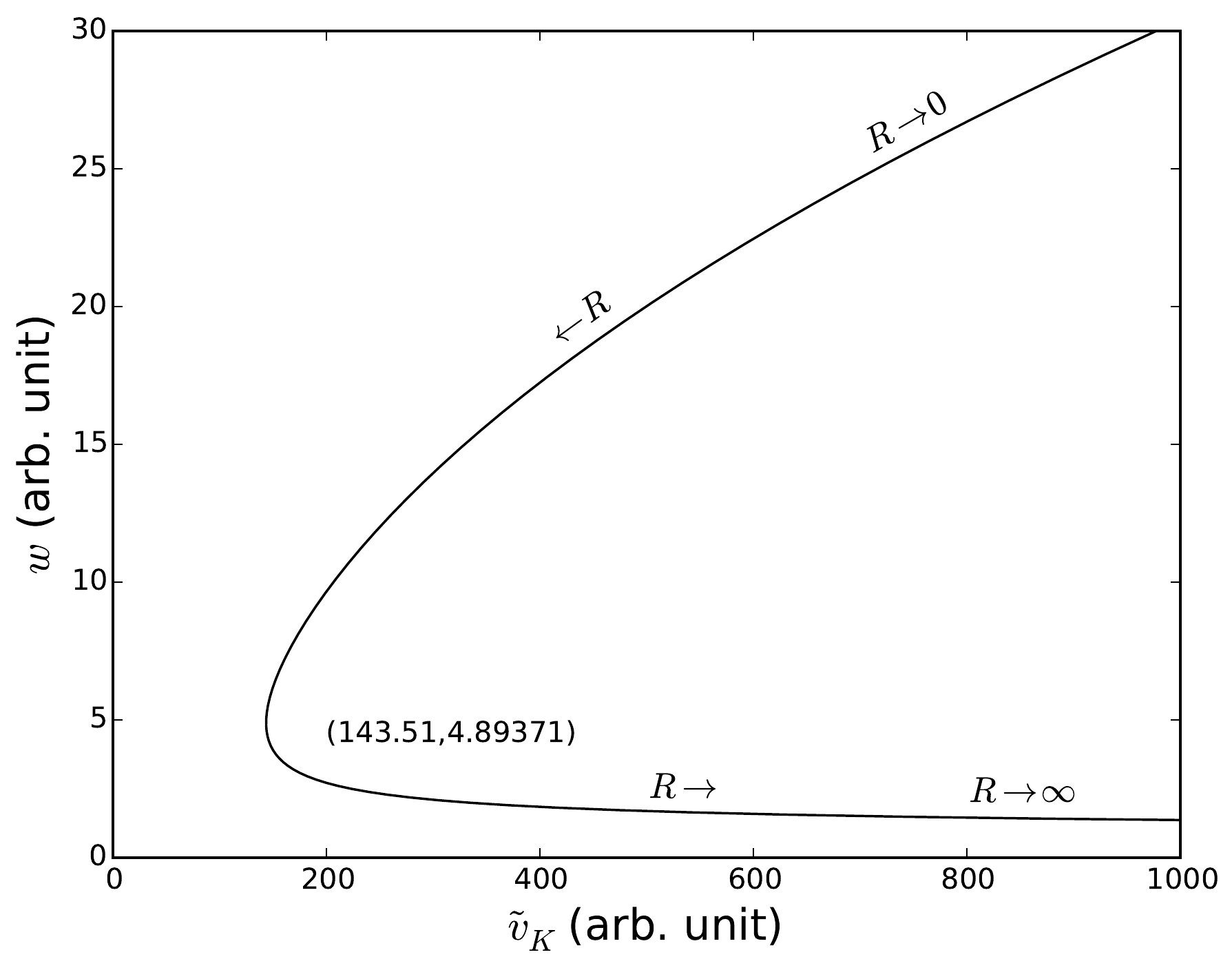}
{\caption*{{\bf Fig. 4}}}
\end{figure}

\newpage
\begin{figure}[h]
\centering
\includegraphics[scale=0.7]{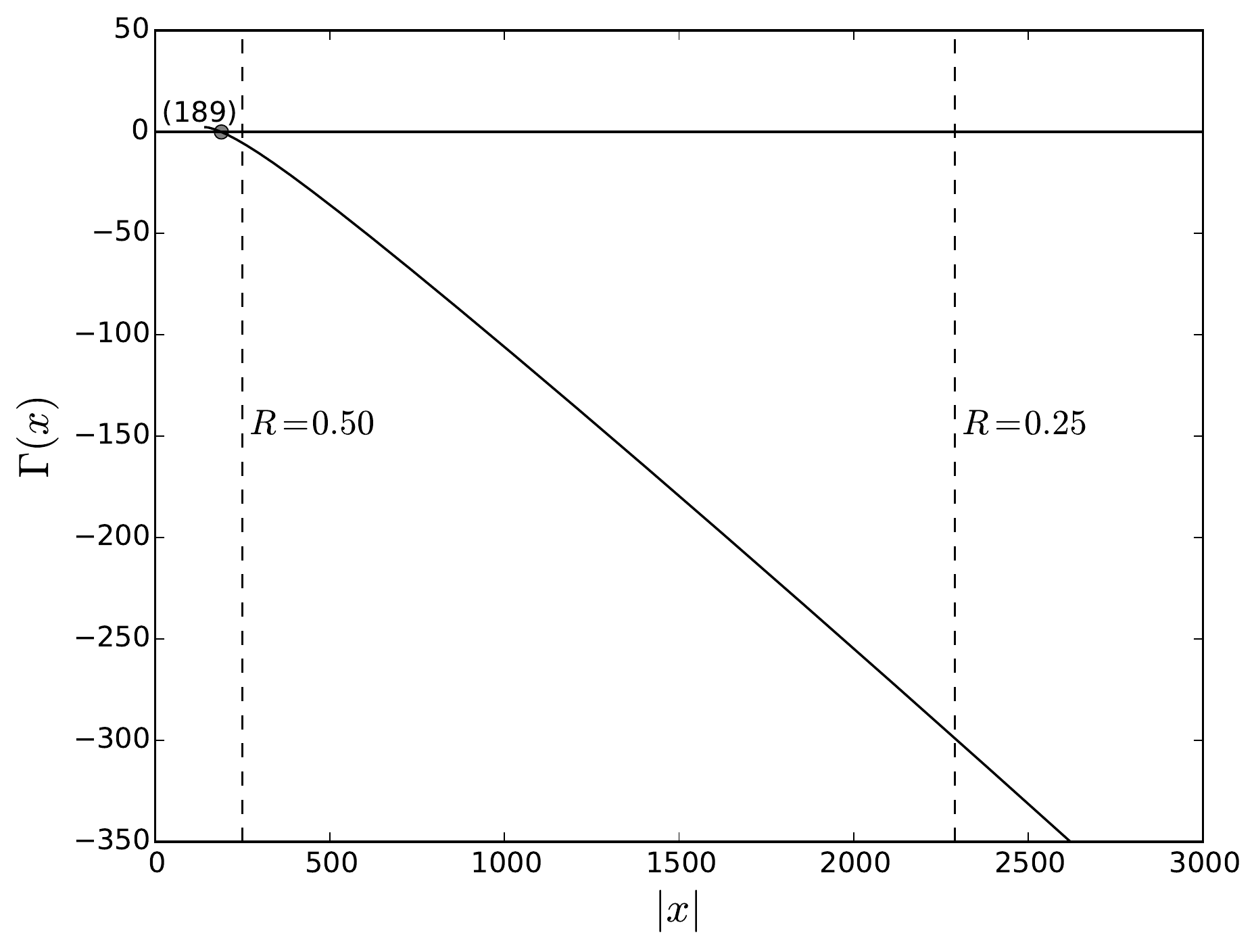}
{\caption*{{\bf Fig. 5}}}
\end{figure}

\newpage
\noindent
{\it keywords:}	graphene, charge density waves, electronic structure, mathematical model, stability.\\

\noindent
PACS, 	73.22Pr (electronic structure of graphene)\\
\indent	\ \ \ 71.45Lr (charge density wave systems)


\newpage
\begin{thebibliography}{9}

\bibitem{Ref_1} P. R. Wallace, “The Band Theory of Graphite”, Phys. Rev.~, {\bf 71}, 622 (1947).

\bibitem{Ref_2} K. S. Novoselov, A. K. Geim, S. V. Morozov, D. Jiang, Y. Zhang, S. V. Dubonos, I. V. Grigorieva, and A. A. Firsov, “Electric field effect in atomically thin carbon films”, Science, {\bf 306}, 666 (2004).

\bibitem{Ref_3} H. Fr\"{o}hlich, “On the theory of superconductivity: the one-dimensional case”, Proc. Roy. Soc. (London) A, {\bf 223}, 296 (1954).

\bibitem{Ref_4} E. Cartlidge, “Graphene superconductivity seen”, Physics World, {\bf 29}, no. 10, 6 (2015). 

\bibitem{Ref_5}  See for example K. G. Rahnejat, A. Howard, N. E. Shuttleworth, S. R. Schofield, K. Iwaya, C. F. Hirijibehedin, C. H. Renner, G. Aeplli and M. Ellerby, “Charge density waves in the graphene sheets of the superconductor CaC6”, Nature Communications, {\bf 2}, article number 558, 29  November (2011).

\bibitem{Ref_6} N. W. Ashcroft and N. D. Mermin, Solid State Physics (W. B. Saunders, Philadelphia, 1976), Chapter 10.

\end{thebibliography}
\end{document}